\documentclass{jpsj-suppl}
\usepackage{txfonts} 
\usepackage{bm}
\usepackage{color,ulem}

\newcommand{\mrm}{\mathrm}
\newcommand{\mcl}{\mathcal}

\title{Lifshitz Transition Induced by Magnetic Field in Frustrated Two-Leg Spin-Ladder Systems}

\author{Takanori \textsc{Sugimoto}$^{1}$, Michiyasu \textsc{Mori}$^{2}$, Takami \textsc{Tohyama}$^{1}$, and Sadamichi \textsc{Maekawa}$^{2}$ }

\inst{$^{1}$ Faculty of Science, Tokyo University of Science, 6-3-1 Niijuku, Katsushika, Tokyo 125-8585, Japan \\
$^{2}$Advanced Science Research Center, Japan Atomic Energy Agency, Tokai, Ibaraki, 319-1195, Japan}

\email{sugimoto.takanori@rs.tus.ac.jp}
\recdate{September 15, 2014}

\abst{The magnetization curve in a frustrated two-leg spin-ladder system is theoretically studied using both analytical and numerical methods. 
This spin system is mapped onto a hard-core boson system, which is composed of bond-operators describing the triplon excitation. 
By the analytical method using the mean-field theory on the bond-operators, we show that cusp singularities emerge in the magnetization curve as a function of a magnetic field due to the strong frustration.
This originates from a change of number of Fermi points in the bosonic dispersion relation with the applied field. It is analogous to {\it Lifshitz transition}.
This singularity is clarified by numerical calculation with the density-matrix renormalization-group method.
Our results will be useful to understand the magnetization process observed in the frustrated two-leg spin-ladder compound BiCu$_2$PO$_6$.}

\kword{Lifshitz transition, cusp singularity, magnetization curve, frustration, spin ladder, BiCu$_2$PO$_6$}

\begin{document}
\maketitle

\section{Introduction}
Recent studies on quantum phase transitions and quantum critical phenomena are mostly concentrated on the low-dimensional physics such as the Mott transition in strongly-correlated electrons and the spin-liquid phase in quantum spin systems.
To clarify their various behaviors, quantum spin systems in low dimensions have been intensively studied as a playground of them.
In order to understand the magnetic behaviors, magnetization process gives essential informations both theoretically and experimentally.
Actually, a pile of previous works has revealed a variety of magnetic phases and quantum phase transitions, e.g., BKT transition, magnetization plateaux, and so on.

Very recently, an experimental study on the magnetization process in BiCu$_2$PO$_6$ has discovered sequential phase transitions induced by magnetic field.\cite{kohama12,kohama14}
According to the report, since the phase transitions in high magnetic fields accompany structural phase transitions, a coupling of lattice degree of freedom is crucial to understand them.
On the other hand, there are also other sequential transitions without structural change in low magnetic fields.
These transitions should be understood in the effective spin model of BiCu$_2$PO$_6$.
Therefore, in this paper, we study the latter sequential phase transitions in low magnetic fields by the effective spin model.

This paper is organized as follows.
In Sec.~2, we show the relationship between the dispersion relation of triplon and magnetization curve.
The bond-operator mean-field approximation~\cite{sachdev90,gopalan94} and the strong rung-coupling limit are used to analyze them.
We also discuss an origin of the magnetic phase transition.
Additionally, we show numerical results obtained by using the density-matrix renormalization-group method~\cite{white} for frustrated two-leg spin ladder in Sec.~3.
The same structure can be found in the numerical results.
Both the analytical and the numerical results are summarized in Sec.~4, and discuss some remaining problems.

\section{Bond-Operator Mean-Field (BOMF) Approximation: Triplon Model}

We consider the following Hamiltonian, which we call frustrated two-leg spin ladder (2LSL), as the effective spin model of BiCu$_2$PO$_6$~\cite{koteswararao07,mentre09,tsirlin10}:
\begin{equation}
\mcl{H}=\mcl{H}_1+\mcl{H}_2+\mcl{H}_\perp+\mcl{H}_{\mrm{Z}}
\label{eq:ham0}
\end{equation}
with
\begin{align}
\mcl{H}_1&=J_1\sum_j\left(\bm{S}_{j,\,\mrm{u}}\cdot\bm{S}_{j+1,\,\mrm{u}}+\bm{S}_{j,\,\mrm{l}}\cdot\bm{S}_{j+1,\,\mrm{l}}\right),\\
\mcl{H}_2&=J_2\sum_j\left(\bm{S}_{j,\,\mrm{u}}\cdot\bm{S}_{j+2,\,\mrm{u}}+\bm{S}_{j,\,\mrm{l}}\cdot\bm{S}_{j+2,\,\mrm{l}}\right),\\
\mcl{H}_\perp&=J_\perp\sum_j\bm{S}_{j,\,\mrm{u}}\cdot\bm{S}_{j,\,\mrm{l}},\\
\mcl{H}_{\mrm{Z}}&= H^z \sum_{j} \left({S^z}_{j,\,\mrm{u}}+{S^z}_{j,\,\mrm{l}}\right)
\end{align}
where $J_1(>0)$ and $J_2(>0)$ are the magnitudes of the antiferromagnetic nearest-neighbor and next-nearest-neighbor exchange interactions, respectively, and $J_\perp(>0)$ is that of the antiferromagnetic nearest-neighbor interaction in the rung direction.
$\bm{S}_{j,\,\mrm{u(l)}}$ is the $S=1/2$ spin operator on the $j$ site in the upper (lower) chain.~\cite{note,casola13}

In this model, there are two possibilities of the ground-state phase without applied magnetic fields: the columnar-dimer and the rung-singlet phases.
The previous works have claimed that the real compound is located in the rung-singlet phase similar to that of non-frustrated 2LSL~\cite{lavarelo12,sugimoto13,plumb13}.
In this phase, two spins on a rung become a singlet pair and the elementary excitation is described by a hard-core boson of a triplet pair on a rung, {\it ``triplon''}.
Low-energy physics and the magnetization process in low magnetic fields are understood by using the triplon picture.
However, the real compound exhibits additional magnetic-phase transitions, which do not exist in the triplon picture without frustrations.
Thus, we can suppose that such a sequential phase transition must be induced by the magnetic frustration.

We consider the strong rung-coupling limit $J_1/J_\perp(\equiv \lambda)\to 0$ with a finite frustration $J_2/J_1(\equiv \eta) \sim 1$.
Since the magnetic behaviors in the rung-singlet phase belong to the same universality class in this limit, the strong rung-coupling limit can be justified qualitatively.
In this limit, the BOMF approximation~\cite{sachdev90,gopalan94,lavarelo12} works well and the low-energy physics can be described by a hard-core boson (triplon).
Then, the Hamiltonian of spin-1/2 operators~(\ref{eq:ham0}) can be rewritten as follows,
\begin{equation}
\mcl{H}_{\mrm{trp}}\cong\mcl{H}_{\mrm{K}}+\mcl{H}_{\mrm{U}}+\mcl{H}_{\mrm{CP}}
\label{eq:ham1}
\end{equation}
with
\begin{align}
\mcl{H}_{\mrm{K}} &= \sum_{q,\alpha} E(q) \, t_{q,\alpha}^\dagger t_{q,\alpha},\\
\mcl{H}_{\mrm{U}} &= \sum_{j,R} \frac{J_R}{2} (m_{j,+}-m_{j,-})(m_{j+R,+}-m_{j+R,-}),\\
\mcl{H}_{\mrm{CP}} &= \sum_{j,R} H^z\, (m_{j,+}-m_{j,-}),
\end{align}
where $\alpha=0,\pm $ and $R=1,2$. 
The creation (annihilation) operator of the triplon is represented as $t_{q,\alpha}^\dagger$ ($t_{q,\alpha}$) in the momentum space. 
These operators obey the statistics of hard-core bosons. 
The number operator of the triplon at $j$-th rung is denoted by $m_{j,\alpha}$.
In this model, magnetic field $H^z$ plays a role of chemical potential for the triplon.
We note that expectation value of subtraction $\langle m_{j,+}-m_{j,-}\rangle$ corresponds to local magnetization at $j$-th rung along the $z$ axis. 

Within the first order of $\lambda$ without the magnetic field, the normalized dispersion relation $\varepsilon(q)\equiv E_q/J_\perp$ is given by,
\begin{align}
\varepsilon(q) \cong 1+\lambda \left(\cos q+\eta \cos 2q\right) = 1+ \lambda \left[2\eta\left(\cos q +\frac{1}{4\eta}\right)^2-\left(\frac{1}{8\eta}+\eta\right)\right].
\end{align}
There is a Lifshitz point at $\eta_{\mrm{L}}=1/4$~\cite{landaubook,fisher80}: the wavenumber of the spin-spin correlations changes from the commensurate ($\pi$) to an incommensurate one ($\neq \pi$) at the Lifshitz point.
Figure 1(a) shows the dispersion relation which has the minimum energy at $q^\ast=\pi$ for $\eta<\eta_{\mrm{L}}$ (commensurate case) and at $q^\ast = \cos^{-1} (-1/4\eta)$ for $\eta\geq \eta_{\mrm{L}}$ (incommensurate case).
The density of states (DOS) dramatically changes at the Lifshitz point $\eta_\mrm{L}$ (Fig.~1(b)):
\begin{equation}
D(\varepsilon)=\begin{cases}
D_-(\varepsilon) & (\eta\leq \eta_L) \\
D_+(\varepsilon)+D_-(\varepsilon) & (\eta> \eta_L) \\
\end{cases}
\end{equation}
with
\begin{equation}
D_\pm(\varepsilon)=\left|\frac{dq}{d\varepsilon} \right|_\pm = \frac{4\eta}{\lambda} \left\{f(\varepsilon) \sqrt{16\eta^2-[1\pm f(\varepsilon)]^2}\right\}^{-1}
\end{equation}
and
\begin{equation}
f(\varepsilon)=\sqrt{1+\frac{8\eta(\varepsilon-1)}{\lambda}+8\eta^2}.
\end{equation}
The divergences of the DOS locate at $\varepsilon_0=1+\lambda(1+\eta)$ and $\varepsilon_\pi=1-\lambda(1-\eta)$ in the commensurate case. On the other hand, there emerges one more divergence at $\varepsilon^\ast=1-\lambda(\eta+1/8\eta)$ in the incommensurate case.

\begin{figure}[tbh]
\centering
\includegraphics[scale=0.5]{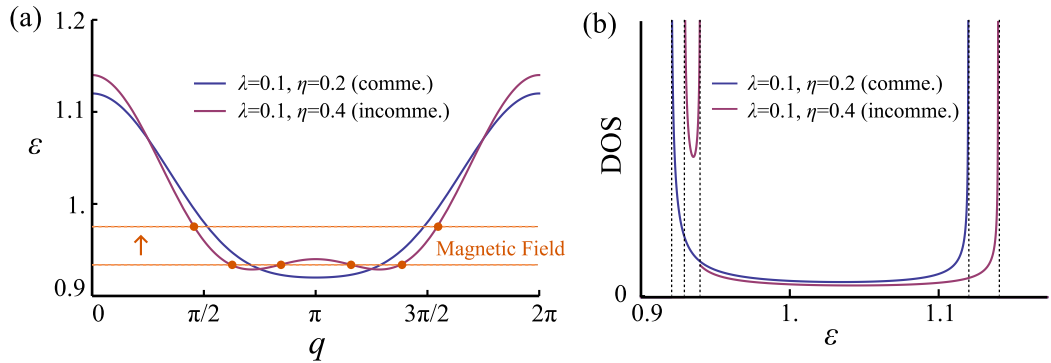}
\caption{Dispersion relation (a) and DOS (b) of triplon for both commensurate and incommensurate cases.
For commensurate case, number of stationary points in the dispersion relation (divergences in DOS) is two.
In contrast, that for incommensurate case is one more than the commensurate case.
In Fig.~(a), dotted lines and closed circles represent the magnetic field $h$ and the Fermi points, respectively.}
\label{fig:dr}
\end{figure}

\begin{figure}[tbh]
\centering
\includegraphics[scale=0.65]{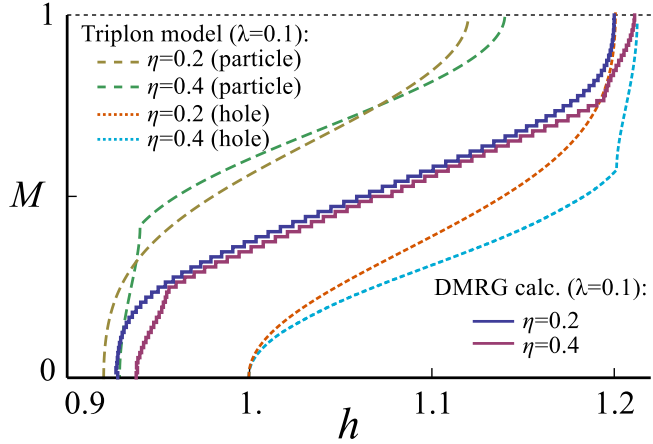}
\caption{Magnetization curve obtained by the triplon model and the DMRG calculation for both commensurate ($\eta=0.2$) and incommensurate ($\eta=0.4$) cases.
In the triplon model, we calculate the magnetization curves by using the dispersion relations of triplon-particle excitation in the non-magnetized ground state (dashed lines), and the triplon-hole excitation in the full-magnetized ground state (dotted lines).
In the magnetization curve obtained by triplon model, a cusp-like singularity emerges for the incommensurate case as compared with the commensurate one.
There is also a similar structure at low magnetic fields in the DMRG calculation (solid lines).}
\label{fig:mh}
\end{figure}

Since the magnetization is the number of the triplons $M=\sum_q m_{q,-}/N$, where $N$ is the number of rungs, the magnetization increases as increasing the chemical potential $h\equiv H^z/J_\perp$. 
If we consider that magnetic field is much less than the saturated field, the repulsive Hamiltonian $\mcl{H}_{\mrm{U}}$ is less effective than the others. 
In this paper, since we consider only the low magnetic field, the repulsive Hamiltonian is neglected.
Within the approximation, we obtain the magnetization in the magnetic field by,
\begin{equation}
M=\frac{1}{\pi}\int_0^{\pi} dq \,\langle \mrm{g.s.}| \,m_{q,-}\, |\mrm{g.s.}\rangle 
\end{equation}
with the ground state,
\begin{equation}
|\mrm{g.s.}\rangle = \Pi_{\{q\,|\,\varepsilon(q)-h<0\}}\, t_{q,-}^\dagger |0\rangle,
\end{equation}
where $|0\rangle$ is the ground state without the magnetic field.
The expectation value of the number of triplons $\langle 0| \,m_{q,-}\, |0\rangle =0$.
There are two phase transitions and three phases in the commensurate case:
\begin{equation}
M(h)=
\begin{cases}
0 &  (h \leq \varepsilon_\pi)  \\
1-\frac{1}{\pi}\cos^{-1}\left( -\frac{1}{4\eta} \left[1-f(h)\right]\right) & (\varepsilon_\pi < h < \varepsilon_0) \\
1 &  (h \geq \varepsilon_0).
\end{cases}
\end{equation}
On the other hand, in the incommensurate case, number of phases is different from the commensurate case:
\begin{equation}
M(h)=
\begin{cases}
0 &  (h \leq \varepsilon_\pi)  \\
\frac{1}{\pi}\left[\cos^{-1}\left( -\frac{1}{4\eta} \left[1+f(h)\right]\right)-\cos^{-1}\left( -\frac{1}{4\eta} \left[1-f(h)\right]\right)\right] & (\varepsilon^\ast < h < \varepsilon_\pi) \\
1-\frac{1}{\pi}\cos^{-1}\left( -\frac{1}{4\eta} \left[1-f(h)\right]\right) & (\varepsilon_\pi < h < \varepsilon_0) \\
1 &  (h \geq \varepsilon_0) .
\end{cases}
\end{equation}
Thus, we can obtain a cusp-like singularity at $h=\varepsilon_\pi$ only in the incommensurate case, that originates from the strong frustraion (see Fig.~\ref{fig:mh}).

Our approximation with triplon particles does not discribe the physics at high magnetic field well, since many-body interactions of triplons are neglected.
On the other hand, we can approach the magnetization curve at a high magnetic field by using the {\it triplon-hole} picture, in which a rung-singlet in the fully-magnetized ground state is associated with a triplon-hole.
With the similar procedure as the triplon-particle picture, we can also obtain the magnetization curves extended from the saturated magnetization.
Figure 2 shows that the magnetization curve in the triplon-hole picture has another cusp with a strong frustration.
Therefore, we expect that the strong frustration induces two cusp singularities at low and high magnetic fields.

This singularity is induced by a change of the number of Fermi points as increasing the magnetic field.
This is very similar to the Lifshitz transition, which is described as a topological change of the Fermi surface for fermion systems.~\cite{lifshitz60}
In addition, this singularity has also been reported for other quantum spin systems: the zigzag spin chain and the frustrated Kondo necklace.~\cite{okunishi99,okunishi01,yamamoto01,yamamoto03} 

\section{Density-Matrix Renormalization-Group (DMRG) Calculation: M-H Curve in 2LSL}
We have also investigated the magnetization process by using the DMRG method. 
The minimum energy $E_m$ in Hilbert subspace of $\sum_{j,i} S_{j,i}^z=-m$ has been calculated for every $m$ with fixed parameters $\lambda$ and $\eta$ without magnetic fields.  
We have performed the calculations for two parameters $\eta=0.2$ (commensurate case) and $\eta=0.4$ (incommensurate case) with fixed $\lambda=0.1$ in a 72-rung ladder.
The truncation errors are less than $10^{-7}$ with the truncation number $300$. 
We checked that the energy converges within an error less than $10^{-5}$.

In an applied field $H^z$, the eigenenergy of the magnetization $m$ linearly decreases as, $E_M-mH^z$.
Thus, the energy levels of the magnetization $m$ and $m+1$ go across at the magnetic field $H_{m:m+1}^z = E_{m+1}-E_m$.
Figure 2 shows that the normalized magnetic field $h=H_{m:m+1}^z/J_p$ for the normalized magnetization $M=m/N$.
We can see the same behavior as the triplon model in the DMRG calculation in low magnetic fields, that is, the cusp singularity emerges with the strong frustration.
In addition, the cusp singularity in high magnetic field is also understood with the dispersion relation of triplon-hole excitation in the full-magnetized ground state.
However, there are a quantitative and a qualitative differences: a difference of the scale of the magnetic field and that of the magnitude of the magnetization at the cusp.
The magnetic field in the triplon model should be rescaled by the repulsive term $\mcl{H}_U$, which is neglected in the analytical approximation.
In addition, the many-body interactions of the repulsive term are not be negligible, if the number of triplons is large.
Thus, we have to deal the change of the dispersion relation of triplons in high magnetic fields, if we discuss the magnitude of the magnetic field at the cusp.

\section{Summary and Discussions}
We have theoretically studied the magnetization curve in the frustrated 2LSL system, which is the effective spin model of BiCu$_2$PO$_6$.
The recent experimental result has shown that there is a sequential magnetic phase transition in the low magnetic field, which has not been clarified so far.
The ground state of the compound is located in the rung-singlet phase, whose qualitative behavior can be understood by using the BOMF approximation. 
The approximation gives the triplon picture, which obeys the hard-core boson statistics and provides a good correspondence with the original spin model in the strong rung-coupling limit.
To clarify the magnetic phase transitions in the low magnetic field in the compound, we have investigated the magnetization process in the strong rung-coupling limit both analytically and numerically.

By using the approximation, we have found that the cusp singularities emerge with a strong frustration.
The singularities originate from the change of the number of Fermi points in the triplon's dispersion relation with applied magnetic fields, which is analogous to the Lifshitz transition.
We have also found the same structure in the numerical results with the DMRG method, although there are two differences: scale of the magnetic field and the magnitude of the magnetization at the cusp.
The scale of the magnetic field and the magnitude of the magnetization at the cusp in the approximation should be modified correctly with the repulsive term in the Hamiltonian.

\section*{Acknowledgements}
This work was partly supported by Grant-in-Aid for Scientific Research (Grant No.23340093, No.24360036, No.24540387, No.25287094, and No.26108716) and bilateral program from MEXT, and by the inter-university cooperative research program of IMR, Tohoku University. 
Numerical computation in this work was carried out on the supercomputers at JAEA and ISSP, The University of Tokyo.



\begin{thebibliography}{99}
\bibitem{note} Our model does not include the alternation in $J_2$, whose effects have been discussed by Refs.~\cite{tsirlin10,casola13}.
\end{thebibliography}
\end{document}